\documentclass[letterpaper]{article} 
\usepackage{aaai25}  
\usepackage{times}  
\usepackage{helvet}  
\usepackage{courier}  
\usepackage[hyphens]{url}  
\usepackage{graphicx} 
\usepackage{natbib}  
\usepackage{caption} 
\usepackage{algorithm}
\usepackage{algorithmic}

\usepackage[inline]{enumitem}
\usepackage{multirow}
\usepackage{color}
\usepackage{multirow}
\usepackage{booktabs}
\usepackage{amsmath}
\usepackage{makecell}

\usepackage{newfloat}
\usepackage{listings}
\DeclareCaptionStyle{ruled}{labelfont=normalfont,labelsep=colon,strut=off} 
\floatstyle{ruled}
\newfloat{listing}{tb}{lst}{}
\floatname{listing}{Listing}
\title{Invisible Filters: Cultural Bias in Hiring Evaluations Using Large\\ Language Models}
\author{
    Pooja S. B. Rao\textsuperscript{\rm 1,2},
    Laxminarayen Nagarajan Venkatesan\textsuperscript{\rm 2}, 
    Mauro Cherubini\textsuperscript{\rm 1},\\
    Dinesh Babu Jayagopi\textsuperscript{\rm 2}
}
\affiliations{
    \textsuperscript{\rm 1}University of Lausanne, Lausanne, Switzerland\\
    \textsuperscript{\rm 2}International Institute of Information Technology Bangalore, Bangalore, India\\
        pooja.rao@unil.ch,
        laxminarayen.nv@iiitb.ac.in,
        mauro.cherubini@unil.ch,
        jdinesh@iiitb.ac.in        
}

\begin{document}

\maketitle

\begin{abstract}
Artificial Intelligence (AI) is increasingly used in hiring, with large language models (LLMs) having the potential to influence or even make hiring decisions.
However, this raises pressing concerns about bias, fairness, and trust, particularly across diverse cultural contexts. Despite their growing role, few studies have systematically examined the potential biases in AI-driven hiring evaluation across cultures.
In this study, we conduct a systematic analysis of how LLMs assess job interviews across cultural and identity dimensions.
Using two datasets of interview transcripts, 100 from UK and 100 from Indian job seekers, we first examine cross-cultural differences in LLM-generated scores for hirability and related traits.
Indian transcripts receive consistently lower scores than UK transcripts, even when they were anonymized, with disparities linked to linguistic features such as sentence complexity and lexical diversity.
We then perform controlled identity substitutions (varying names by gender, caste, and region) within the Indian dataset to test for name-based bias. These substitutions do not yield statistically significant effects, indicating that names alone, when isolated from other contextual signals, may not influence LLM evaluations.
Our findings underscore the importance of evaluating both linguistic and social dimensions in LLM-driven evaluations and highlight the need for culturally sensitive design and accountability in AI-assisted hiring.


\end{abstract}

%

\section{Introduction}



Artificial intelligence (AI) has rapidly transformed the hiring landscape, reshaping how employers source, screen, and evaluate candidates \cite{oecdOECDEmploymentOutlook2023}.  A recent survey by Harvard Business School \cite{fullerHiddenWorkersUntapped2021} highlights the widespread adoption of AI-enabled recruiting technologies: approximately 63\% of employers reported using a Recruitment Management System (RMS), a figure that rises to 69\% for organizations with over 1,000 employees. In the U.S., this number is even higher at 75\%, compared to 58\% in the U.K. and 54\% in Germany. These systems are not limited to administrative support. They are widely used to filter and rank candidates, with over 90\% of employers employing them for initial screening.

Job interviews remain the most common selection method for employment \cite{trossEffectCoachingInterviewees2008}, and digital platforms are being increasingly used for hiring \cite{european2018automation}.
Asynchronous video interviews (AVI), also called online video interviews \cite{Langer2017}, allow candidates to complete interviews at their convenience and are gaining popularity due to their scalability and ease of use. 
Many employers automate many aspects of the hiring process, from candidate scoring to interview scheduling. 
AVIs \cite{raoAutomaticAssessmentCommunication2017, rasipuramAsynchronousVideoInterviews2016} and video resumes \cite{nguyenHirabilityWildAnalysis2016} are used in early screening rounds\footnote{For example, \url{www.hirevue.com}, \url{www.modernhire.com}, \url{www.viasto.de}. Last accessed August 2025}, leveraging AI techniques \cite{chenEthicsDiscriminationArtificial2023} like computer vision, natural language processing, and audio processing to score candidates.
As AI capabilities have expanded, large language models (LLMs) like GPT-4 \cite{openaiGPT4TechnicalReport2024} and Gemini \cite{teamGeminiFamilyHighly2025} have emerged as particularly powerful tools in these pipelines. Their accessibility through APIs, user-friendly interfaces, and support for rapid prototyping \cite{wangFarsightFosteringResponsible2024} makes them especially appealing for companies that may have previously avoided AI tools due to cost or technical barriers. As a result, the integration of LLMs into hiring workflows is not only feasible—it is likely inevitable.

Yet, as AI and LLMs in particular assume increasingly central roles in hiring decisions, there are urgent concerns about fairness, accountability, and cultural bias. Previous studies have demonstrated that text-to-image models \cite{qadriAIsRegimesRepresentation2023, ghoshGenerativeAIModels2024} and LLMs often encode and prioritize Western norms and values \cite{agarwalAISuggestionsHomogenize2025}, leading to representational harms for non-Western users and contexts. These issues are especially pronounced in global South settings, such as South Asia, where Western-centric models can marginalize culturally specific forms of expression and communication \cite{qadriAIsRegimesRepresentation2023}.
Previous work has shown discriminatory hiring practices based on gender and race \cite{chenEthicsDiscriminationArtificial2023}, including those using LLMs for resume screening tools \cite{wilsonGenderRaceIntersectional2025}.
While explicit forms of stereotyping in AI outputs are concerning, more subtle and insidious biases pose an equally serious threat. Recent research shows that LLMs can shape user behaviour through autocomplete and writing suggestions, altering not just syntax but also tone, content, and even attitudes toward social and cultural issues \cite{hohensteinArtificialIntelligenceCommunication2023, poddarAIWritingAssistants2023, agarwalAISuggestionsHomogenize2025}. This raises a critical, yet underexplored question: do such cultural biases extend to how LLMs evaluate texts in high-risk contexts, such as job interviews?

The field of fairness and ethics has increasingly called for a rethinking of how AI harms are conceptualized and studied. Scholars have emphasized the need for community-centered approaches and for frameworks that move beyond Western-centric notions of fairness. However, this remains an under-addressed issue, as research into ethics and fairness often centres Western perspectives and relies on Western frameworks of ethics and fairness \cite{septiandriWEIRDFAccTsHow2023, narayananvenkitNationalityBiasText2023}. This shift toward more inclusive and context-sensitive evaluation is essential if we are to understand how AI systems behave across diverse cultural and social settings, particularly in high-stakes domains like employment.

This paper makes three key contributions to understanding cross-cultural variation and potential bias in LLM-based evaluations of job interview transcripts.
\begin{enumerate}
    \item \textit{Cross-Cultural Comparison of LLM Evaluations:} We analyze two corpora of interview transcripts—100 from UK-based job seekers and 100 from Indian job seekers—to examine how LLMs assess responses across Western and non-Western contexts. By anonymizing identity-related entities, we isolate linguistic and semantic features as the primary source of variation. We evaluate each transcript on job attributes like hireability, positive impression, self-promotion and use of stories, considered important in an interview, using two widely popular LLMs, GPT-4o and Gemini Flash1.5. Results show that Indian transcripts consistently receive lower scores than UK transcripts across all evaluated dimensions, raising concerns about systematic scoring disparities.
    \item \textit{Linguistic Feature Analysis:} To understand the basis of these scoring disparities, we extract a range of linguistic features like type-to-token ratio, sentence length, confidence markers (e.g., hedging vs assertive language) and examine their correlation with LLM-generated scores. Our findings indicate that lexical diversity, syntactic complexity, and readability strongly influence evaluations. Importantly, even after controlling for these features, UK transcripts continue to score higher, suggesting culturally conditioned preferences embedded in LLM behaviour.
    \item \textit{Controlled Identity Substitution to Probe Bias}: Previous work has shown that popular LLMs show bias towards marginalized Indian groups \cite{khandelwalIndianBhEDDatasetMeasuring2024}. We further explore whether identity-based cues influence LLM evaluations by systematically altering names within the Indian dataset to signal gender, caste, and regional identity, while holding all other content constant. These controlled substitutions reveal no statistically significant differences in LLM scores, suggesting that names alone, in the absence of additional contextual cues, may not be sufficient to trigger bias. These findings offer a cautiously optimistic signal in the ongoing discussions on fairness and highlight the utility of controlled counterfactual designs for bias testing.
\end{enumerate}

It is important to clarify that this study is not intended as a comprehensive evaluation of LLM capabilities in hiring contexts. We do not present our findings as definitive assessments of any specific model or of LLMs in general. Rather, this targeted investigation serves as an illustrative exploration of how identity cues and linguistic differences across cultures may influence AI-generated evaluations. Despite its limitations, the study highlights critical considerations for the design and deployment of LLMs in high-stakes domains like hiring and invites deeper reflection on fairness, transparency, and the risks of AI in critical decision-making.

\section{Related Work}

We review scholarly work examining cultural bias in AI systems, particularly how large language models (LLMs) prioritize Western linguistic norms, values, and communication styles. 
Given our focus on AI-assisted evaluation of written job interview responses, we also review literature on the evaluation of hiring technologies powered by LLMs. This intersection of sociotechnical systems, language, and cross-cultural evaluation frames the core concerns of our work: how seemingly neutral AI recommendations may carry invisible cultural filters that influence high-stakes decision-making processes such as hiring.

\subsection{Cultural Bias in LLMs}

Large language models are known to encode cultural and linguistic biases present in their training data. Because most LLMs are trained on English corpora dominated by Western sources, they tend to favour Western norms and values in their outputs \cite{johnsonGhostMachineHas2022,atariWhichHumans2023}, primarily when prompted in English \cite{caoAssessingCrossCulturalAlignment2023}. For instance, early analyses revealed that GPT-3 exhibited severe religious and ethnic biases, such as disproportionately associating Muslims with violence \cite{abid2021large}. 

Recent studies using cross-cultural benchmarks have confirmed systemic Western alignment in LLM outputs. \citet{tao2023cultural} showed that state-of-the-art LLMs align more closely with Western values (as measured by World Values Survey data) than non-Western ones. \citet{naous2025origin} attribute these patterns to pre-training data imbalances and tokenization mismatches, noting that even multilingual LLMs favour Western-associated entities when operating across languages.
\citet{khandelwalIndianBhEDDatasetMeasuring2024} show that when prompted with Indian context data, popular LLMs exhibit a strong propensity to produce stereotyped outputs about marginalized Indian groups (e.g., caste and religion), far more than the bias observed on traditionally studied Western axes like gender or race.

LLMs also tend to homogenize culturally distinct expressions. \citet{agarwalAISuggestionsHomogenize2025} found that AI writing assistants encouraged Indian English speakers to adopt U.S.-centric phrasings, diminishing local nuance. Similarly, \citet{mukherjee-etal-2023-global} expanded bias audits to 24 languages, revealing that social biases are not globally uniform: bias in models can manifest differently in, say, Indian languages versus European languages, and along dimensions like toxicity or ableism that extend beyond the gender/race categories typically examined. 

Outside text, cultural erasure extends to visual generative models. ~\citet{qadriAIsRegimesRepresentation2023} found that text-to-image models presented South Asian subjects through a Western, stereotypical gaze. ~\citet{ghoshGenerativeAIModels2024} echoed this finding, showing generative AI produced exoticized and inauthentic visuals when rendering non-Western prompts.
These problems are particularly evident in global South contexts, where Western-centric frameworks often marginalize culturally specific forms of expression and communication.

Researchers have also explored bias tied to dialect and intersectionality. \citet{deas-etal-2023-evaluation} evaluated model performance on African American English (AAL), finding significant drops in generation quality compared to standard English. \citet{sandovalMyLLMMight2025} noted that Black users felt current LLMs failed to capture the nuance of AAL and preferred systems capable of respectful register-shifting. \citet{maIntersectionalStereotypesLarge2023} further demonstrated that LLMs reproduce intersectional stereotypes (e.g., combining religion, race, and gender) despite safety training. 

These nuanced results show that LLMs can reproduce societal stereotypes targeting complex, intersecting identities, not just single demographics. Prior research establishes a growing awareness that cultural and linguistic biases are \textit{baked into} LLM behaviour. Our work extends this literature by examining such biases in an AI-driven hiring evaluation setting to examine the influence of cross-cultural subtleties (like a candidate’s style or cultural background) in LLM judgments.

\subsection{Bias in Hiring with AI}

Algorithmic hiring systems have repeatedly been shown to replicate existing human and institutional biases. 
\citet{raghavanMitigatingBiasAlgorithmic2020} reviewed commercial hiring algorithms and found that bias-mitigation claims were often unsubstantiated, as vendors frequently failed to disclose meaningful technical documentation, offered no empirical evidence of fairness improvements, and provided vague or inconsistent definitions of bias. Their analysis revealed that many systems marketed as \textit{fair} relied on proprietary scoring mechanisms and opaque inputs, making it difficult for external auditors or employers to assess the legitimacy of fairness claims. 
\citet{chenEthicsDiscriminationArtificial2023} similarly observed that many recruitment platforms inadvertently inherit biases from training data and developer assumptions.
\citet{hangartnerMonitoringHiringDiscrimination2021} developed a cost-efficient methodology to monitor hiring discrimination on online recruitment platforms and found that application callbacks are lower for immigrant and minority ethnic groups.

Numerous real-world incidents have documented the risks of algorithmic bias in hiring: from reinforcing historical inequalities\footnote{\url{https://incidentdatabase.ai/cite/135/}} to employee terminations based on automated assessments\footnote{\url{https://incidentdatabase.ai/cite/111/}} \cite{mcgregorPreventingRepeatedReal2021}.  
One infamous example is Amazon’s discontinued hiring model, which penalized resumes containing terms like “women’s chess club” due to training on male-dominated hiring data.\footnote{\url{https://incidentdatabase.ai/cite/37/}} 
Traditional correspondence tests, where researchers send fictitious but identical CVs except for the minority trait to be tested (for example, names that are deemed to sound ‘Black’ versus those deemed to sound ‘white’), have quantified discrimination in hiring practices \cite{bertrandAreEmilyGreg2004}.
Extending these concerns to AI systems, \citet{wilsonGenderRaceIntersectional2025} found that LLM-based resume screening disadvantaged Black and female-associated names, even when all other resume content was identical. 
Similarly, \citet{puutioFirstComeFirst2025} ran 2,000 simulated applications through ChatGPT and found emergent patterns of discrimination, such as preference for elite education or positional bias in response order.

To address these issues, researchers have proposed a range of mitigation strategies. 
Specific to bias in hiring \citet{albaroudi2024comprehensive} present a taxonomy of fairness techniques, including de-biasing embeddings and preprocessing structured data. \citet{fabris2025fairness} advocate for a multidisciplinary approach that incorporates legal and ethical oversight alongside technical solutions. 
More broadly, the literature outlines strategies ranging from practical interventions to reduce societal harms \cite{kumarLanguageGenerationModels2023,cuiRiskTaxonomyMitigation2024} to frameworks for identifying and documenting risks and biases \cite{raoRiskRAGDataDrivenSolution2025, wangFarsightFosteringResponsible2024, derczynskiAssessingLanguageModel2023}, with the goal of improving practitioner awareness and guiding responsible system design.


\subsection{Research Gap} 
While prior research on AI hiring bias has mostly focused on race and gender within a single-cultural context (often the U.S.), we extend the inquiry to cross-cultural disparities between Indian and UK candidates.
This study builds on the cultural bias findings from the NLP domain and the fairness concerns in algorithmic hiring. By bringing these threads together, we aim to shed light on whether candidates from non-Western or non-mainstream backgrounds face an invisible disadvantage when assessed by LLMs. In summary, our related work review indicates that (1) LLMs carry latent cultural biases favoring specific values, dialects, and demographic stereotypes, and (2) uncritical use of AI in hiring has led to measurable discrimination. These two insights directly motivate our study: how LLM-generated hiring evaluations might inadvertently prefer some cultural-linguistic style over others, and in doing so, our findings contribute to the broader effort of ensuring fairness and cultural equity in AI-supported hiring practices.

\renewcommand{\arraystretch}{1.2} 
\begin{table*}[!htbp]
\caption{\textbf{Metrics used for LLM-based hiring evaluation:} Each metric, its description, and its source from literature are shown. This, with corresponding items, was provided to the LLM in the prompt alongside the interview transcript.
}
\centering
\footnotesize
\label{tab:metrics}
\begin{tabular}{@{}llll@{}}
\toprule
\textbf{Metrics}                                                                 & \textbf{Source questionnaire}                                                                                                                                          & \textbf{Items (N)} & \textbf{Description}                                                                                                                                                                                                                                                                                     \\ \midrule
\textit{\textbf{Hireability}}                                                    & \begin{tabular}[c]{@{}l@{}} \citet{maderaGenderLettersRecommendation2009}\end{tabular}                 & 3                     & \begin{tabular}[c]{@{}l@{}}Likelihood of an individual getting hired, measures the potential \\ that a job seeker will be an effective employee\\ \cite{maderaGenderLettersRecommendation2009}. \end{tabular}                                                                              \\  \hline 
\textit{\textbf{\begin{tabular}[c]{@{}l@{}}Positive \\ impression\end{tabular}}} & \citet{raoPotentialSupportingAutonomy2024}                                                                                                                                           & 3                     & \begin{tabular}[c]{@{}l@{}}Lasting impact an individual has after the interview,  influenced \\by non-verbal behavior {}\cite{imadaInfluenceNonverbalCommunication1978}.\end{tabular}                                                                                                     \\  \hline 
\textit{\textbf{Self-promotion}}                                                 & \begin{tabular}[c]{@{}l@{}} \citet{bolinoMeasuringImpressionManagement1999}\end{tabular} & 4                     & \begin{tabular}[c]{@{}l@{}}Impression management tactics; refers to an attempt to evoke \\ favourable character  perceptions about oneself to elicit respect for \\ one's abilities rather than just being liked \\ \cite{godfreySelfpromotionNotIngratiating1986, Waung2015}.\end{tabular} \\  \hline 
\textit{\textbf{Storytelling}} & \citet{raoPotentialSupportingAutonomy2024}  & 1                     & \begin{tabular}[c]{@{}l@{}}Answering past behaviour questions with a narrative, providing \\information on the situation, task, action and results \\{}\cite{bangerterStorytellingSelectionInterview2014}.\end{tabular}                                                                     \\ \bottomrule
\end{tabular}
\end{table*}

\section{Method}
To examine how LLMs evaluate job interview performance across cultural and identity contexts, we collected and transcribed video interviews from Indian and UK job seekers. The resulting transcripts were analyzed under two contexts: (i) anonymized evaluations and (ii) identity-modified evaluations. 

\subsection{Participant Recruitment and Data Collection}
A priori power analysis indicated that a minimum of 143 participants was required to detect a medium effect size ($f = 0.40$) with $\alpha = 0.05$ and $power=0.95$. To account for potential technical issues during recording, we recruited a total of 200 participants. Screening criteria included:
\begin{enumerate*}[label=(\roman*)]
    \item actively seeking employment;
    \item unemployed, partly employed, or not in paid work;
    \item at least 18 years of age;
    \item willingness to record a video interview;
    \item access to a webcam.
\end{enumerate*} 

\subsubsection{Indian Data}
We recruited 100 participants from our university, of whom 42 were female and 58 were male. All participants were Indian nationals in the age group of 18-23. All participants were students and reported professional fluency in English. 

\subsubsection{UK Data}
We recruited 100 participants via the Prolific platform\footnote{\url{https://www.prolific.co/}}. 
We restricted the participants to those who resided in the United Kingdom and Ireland, with English as their first language.
The sample included 64 females and 36 males. Age distribution was: 18–29 ($n=50$), 30–39 ($n=31$), 40–49 ($n=14$), 50–59 ($n=3$), and 60+ ($n=2$). Their nationalities included UK ($n=93$), Ireland ($n=4$), and the US ($n=3$). 64 participants were students.

\subsubsection{Procedure}
Participants completed asynchronous video interviews using a custom web-based platform. Each session included five questions: one introductory prompt (e.g., “Tell me about yourself”) and four additional questions assessing the interviewee’s competence and knowledge. 
These questions were randomly drawn from a pool spanning six commonly used structured interview domains: work experience, leadership, personality, career goals, skills, and behavioural/situational responses \cite{Wiersma2016, hartwellAreWeAsking2019}. 
We simulated a first-round behavioral job interview for a generic professional role, where no specific job title or responsibilities were included, to mirror common HR screening practices and allow for consistent, role-agnostic evaluation of candidate responses. 
Each response was limited to a maximum of five minutes. 
All videos were transcribed using the OpenAI Whisper API base model \cite{radfordRobustSpeechRecognition2022}. 

\subsection{LLM-Based Hiring Evaluations}

\subsubsection{Metrics}
We assessed candidate performance on four job-relevant attributes identified in personnel selection research: \textit{hireability}, \textit{positive impression}, \textit{self-promotion}, and \textit{storytelling}.
These attributes were selected for their relevance to hiring decisions based on personnel selection literature \cite{huffcuttIdentificationMetaAnalyticAssessment2001} and for being suitable for text-based evaluation.
We used standardized scales to measure each of them, with each attribute having multiple item descriptors scored on a five-point Likert scale (see Table~\ref{tab:metrics} for details). 

\subsubsection{LLM Scoring}
All transcripts were anonymized by replacing identifiable entities (e.g., names, cities) with generic placeholders (e.g., Person X, City Y). We crafted and iteratively refined a scoring prompt instructing the LLM to rate each transcript against four job-relevant metrics. 
The prompt included the definitions of the metrics, corresponding items (see Table \ref{tab:cross_culture}) and their respective scale from 1 to 5. The LLM was asked to embody the role of a recruiter, making hiring decisions without speculating on the identity or background of the candidate. 
Like the data collection process, the prompt simulated a first-round behavioral interview for a generic role, intended to screen candidates for subsequent, job-specific interview stages. 
Each transcript was independently scored by two LLMs: OpenAI’s GPT-4o and Google’s Gemini Flash 1.5. Final scores for each attribute were computed as the mean of its corresponding items.

\subsubsection{Linguistic Features}
To investigate whether linguistic characteristics contributed to observed cross-cultural differences in LLM scores, we extracted several standard textual features known to influence text-based assessments.

\begin{figure*}[!htbp]
    \centering
    \includegraphics[width=0.95\linewidth]{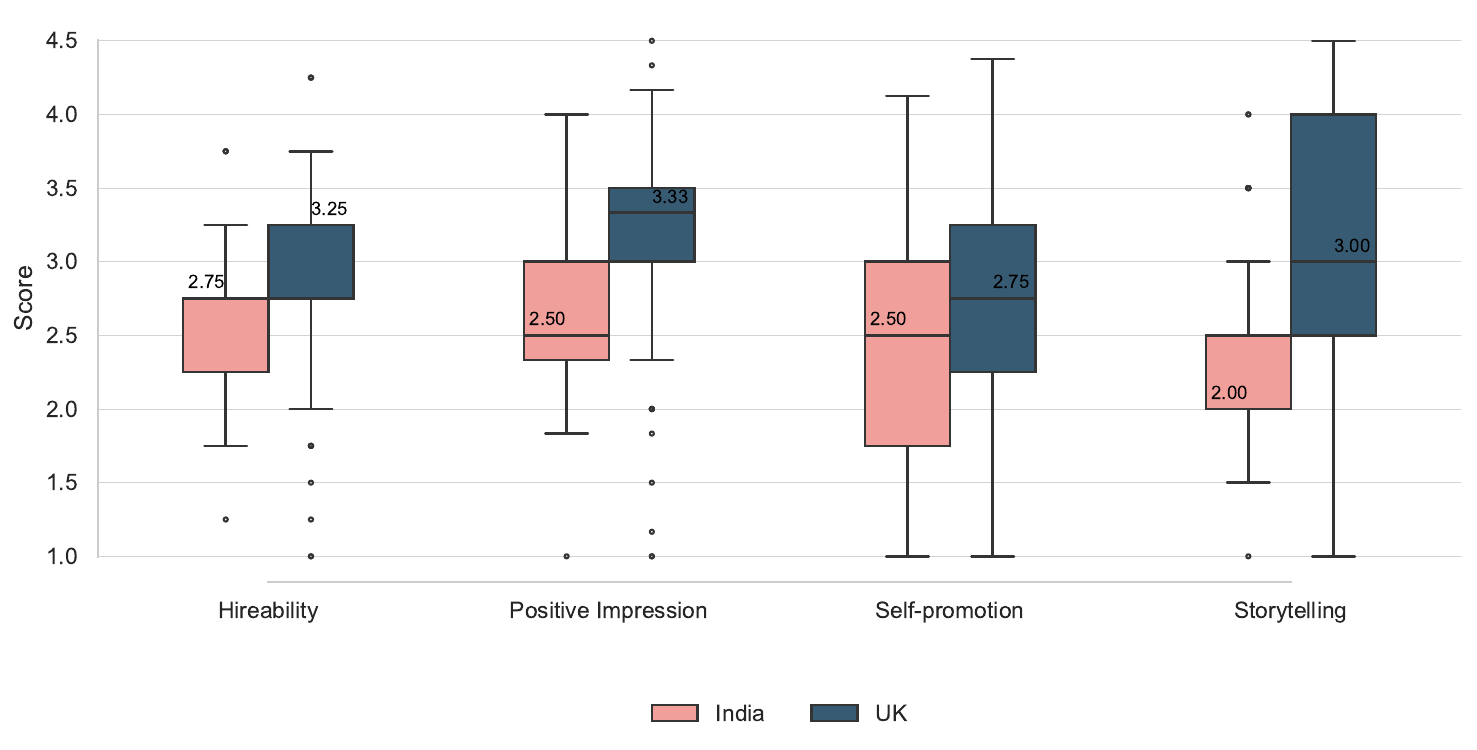}
    \caption{Box plots showing scores for four job-relevant metrics as scored by GPT-4o. LLM scores are significantly lower for Indian transcripts than the UK across all four metrics, indicating a potential bias favouring Western linguistic styles.}
    \label{fig:plot_cross_culture}
\end{figure*}

\begin{enumerate}
    \item \textit{Lexical diversity} was measured using the type-token ratio (TTR), defined as the number of unique words divided by the total number of words, indicating vocabulary diversity.
    \item \textit{Sentence complexity} was operationalized using average sentence length (ASL), computed as total word count divided by the number of sentences. 
    \item \textit{Readability} was quantified using the Flesch Reading Ease (FRE) score \cite{flesch1948new}, calculated as:
    \[
    \textstyle \text{FRE} = 206.835 - (1.015 \times \text{ASL}) - (84.6 \times \text{ASW})
    \] where ASL is the average sentence length and ASW is the average number of syllables per word. Higher scores indicate easier readability.
    \item \textit{Confidence markers} were quantified by calculating the frequency of hedging terms (e.g., “maybe,” “perhaps”) and assertive terms (e.g., “clearly,” “certainly”), expressed as ratios relative to total word count. These measures reflect the speaker's projected level of confidence in their responses \cite{hyland2011metadiscourse}.

\end{enumerate}







\subsubsection{Identity-Based Substitutions}
To assess the influence of identity cues, we performed controlled substitutions of candidate names within the Indian dataset. Using a within-subjects design, we generated identity-modified versions of each transcript by varying candidate names along three dimensions: gender (male/female), region (North/South Indian), and caste (upper/lower), informed by culturally salient naming conventions \cite{embletonNamesIndiaHistory2023}. We chose 5-10 names per combination (for e.g., 10 names for a North Indian upper caste female), leading to a total of 60 names. For each transcript, we substituted a randomly selected name from each combination. All other content remained identical, and anonymised entities except names were unchanged. This produced multiple identity variants per transcript (e.g., the same content with a North Indian upper-caste male name vs. a South Indian lower-caste female name).
To reduce environmental impact and computational cost, we randomly selected 50 transcripts from the Indian dataset, resulting in 400 data points (50 transcripts × 8 identity conditions) and evaluated only using GPT-4o.
This enabled evaluation of LLM sensitivity to socio-cultural identity signals embedded in names.

\setlength{\tabcolsep}{12pt} 
\renewcommand{\arraystretch}{1.1} 
\begin{table*}[!tbp]
\centering
\footnotesize
\caption{\textbf{Statistical significance of the results for cross-cultural comparison:} Mann-Whitney U test results between the Indian and UK LLM-generated scores. The test showed statistically significant differences across all the chosen job-relevant metrics with GPT-4o, except for self-promotion with Gemini Flash 1.5, indicating scoring disparities favouring Western-style responses. Overall final scores for each metric were computed as the mean of its corresponding items. $^{**}$p$<$0.01; $^{***}$p$<$0.001; $ns$:$p>0.05$.}
\label{tab:cross_culture}
\begin{tabular}{@{}lcccc@{}}
\toprule
\multirow{2}{*}{\textbf{Metric}}                                                                                 & \multicolumn{2}{c}{\textbf{GPT-4o}} & \multicolumn{2}{c}{\textbf{Gemini Flash 1.5}} \\ \cmidrule(l){2-5} 
                                                                                                                 & statistic       & p-value           & statistic            & p-value                \\ \midrule
Overall Hireability                                                                                              & 2284.5          & \textbf{***}      & 1888.5               & \textbf{***}           \\
As a recruiter, I would be willing to hire him/her? (H1)                                                         & 2314.0          & \textbf{***}      & 2005.5               & \textbf{***}           \\
As a recruiter, I would perceive him/her as a “top-notch” candidate. (H2)                                        & 2261.0          & \textbf{***}      & 1932.0               & \textbf{***}           \\
As a recruiter, I would think he/she would make an effective employee. (H3)                                      & 2295.0          & \textbf{***}      & 1993.5               & \textbf{***}           \\
As a recruiter, I would consider him/her as an  “excellent” asset. (H4)                                          & 2280.0          & \textbf{***}      & 2096.0               & \textbf{***}           \\ \midrule
Overall Positive Impression                                                                                      & 2168.5          & \textbf{***}      & 1913.5               & \textbf{***}           \\
The performance during the interview was good. (P1)                                                              & 2268.5          & \textbf{***}      & 2015.5               & \textbf{***}           \\
The interviewee gave a good impression of him/herself. (P2)                                                      & 2233.5          & \textbf{***}      & 1848.5               & \textbf{***}           \\
The interviewee looked competent. (P3)                                                                           & 2316.5          & \textbf{***}      & 2585.5               & \textbf{***}           \\ \midrule
Overall Self-Promotion                                                                                           & 3646.5          & \textbf{**}       & 4406.0               & ns                     \\
Talk proudly about their experience or education. (S1)                                                           & 4078.5          & ns                & 5000.0               & ns                     \\
Make people aware of their talents or qualifications. (S2)                                                       & 3619.5          & \textbf{**}       & 4389.5               & ns                     \\
Let others know that their are valuable to the organization. (S3)                                                & 3329.0          & \textbf{***}      & 4036.0               & ns                     \\
Make people aware of their accomplishments. (S4)                                                                 & 3582.5          & \textbf{**}       & 4220.0               & ns                     \\ \midrule
\begin{tabular}[c]{@{}l@{}}Storytelling\\ The interviewee provided a clear and concise story. (ST1)\end{tabular} & 1974.0          & \textbf{***}      & 2346.5               & \textbf{***}           \\ \bottomrule
\end{tabular}
\end{table*}

\subsubsection{Analysis}
To evaluate cross-cultural differences in LLM-generated scores, we compared the outputs for the anonymized Indian and UK transcripts. Due to technical issues in some recordings, we analyzed 99 Indian and 95 UK transcripts. We used the Shapiro–Wilk test to confirm that the score distributions deviated from normality. Consequently, we applied the non-parametric Mann–Whitney U test to assess differences in LLM-assigned scores between the two groups. This analysis was conducted independently for each LLM (GPT-4o and Gemini). 

To further investigate whether linguistic properties contributed to observed scoring disparities, we conducted a post-hoc multiple linear regression analysis using the combined dataset of Indian and UK transcripts. 
The dependent variables were the four job job-relevant metrics.
The predictors included region (India vs. UK) along with linguistic features: type-token ratio, average sentence length, Flesch Reading Ease (FRE), hedging ratio, and assertive ratio. 
This model enabled us to determine whether region remained a significant predictor of scores after controlling for textual features.
Prior to fitting the model, we checked standard regression assumptions. Residual diagnostics indicated mild non-linearity and deviation from normality, confirmed by a significant Shapiro-Wilk test ($p=0.0003$). 
Multicollinearity was assessed using VIFs; most predictors were well below conventional thresholds, though average sentence length and FRE showed high correlation ($r=-0.93$). Both were retained given their relevance and contribution to model fit.


To examine potential identity-based bias within the Indian dataset, we analyzed score changes across identity-substituted variants. A separate linear mixed-effect regression (LMER) model was fitted for each of the four job-relevant metrics with gender, region, and caste (based on name cues) as fixed effects and transcript ID as a random effect to account for repeated measures across the eight identity conditions. This allowed us to test whether socio-cultural identity signals embedded in names influenced LLM evaluations, independent of content.

\section{Results}

\subsection{Cross-Cultural Differences in LLM Evaluations}

We compared the anonymized interview transcripts from Indian and UK participants to assess whether LLMs exhibit differential scoring across cultural contexts. Mann–Whitney U tests were conducted separately for GPT-4o and Gemini Flash 1.5 on each of the four job-relevant metrics—hireability, positive impression, self-promotion, and storytelling—and their component items.

For both LLMs, Indian transcripts consistently received significantly lower scores (Figure \ref{fig:plot_cross_culture}) than UK transcripts on hireability, positive impression, and storytelling, with all $p$-values well below 0.001, indicating strong evidence of a scoring gap. Refer to Table \ref{tab:cross_culture} for details.
Notably, all the corresponding items (H1-H4, P1-P3, ST1) for the three metrics were scored consistently lower. The median scores ranged between 2 to 2.75 for Indian transcripts whereas for UK transcripts it ranged between 2.75 to 3.25 indicating a large gap.
These findings suggest a potential structural bias in LLM evaluations favouring UK linguistic norms or response styles.
In the case of self-promotion, GPT-4o showed significant differences for the overall metric and most sub-items (S2, S3, S4), while Gemini Flash did not show statistically significant differences on any self-promotion measure. 


\subsection{Influence of Linguistic Features}

\setlength{\tabcolsep}{12pt} 

\begin{table*}[!htbp]
\centering
\footnotesize
\caption{\textbf{Regression models for linguistic feature analysis}: Coefficients of the regression models built with linguistic features predicting the job-relevant metrics: hireablity, positive impression, self-promotion and storytelling. Each predictor has its coefficient on the first line and the t-statistic on the line below. The last row indicates Adjusted $R^2$ and F-statistics of the model with degrees of freedom. $^{*}$p$<$0.05; $^{**}$p$<$0.01; $^{***}$p$<$0.001 }
\label{tab:regression_summary}
\begin{tabular}{lcccc}
\toprule
\textbf{Predictor} & \textbf{Hireability} & \textbf{Positive Impression} & \textbf{Self-Promotion} & \textbf{Storytelling} \\
\midrule
Intercept & \makecell{4.218$^{***}$ \\ t = 9.58} & \makecell{4.300$^{***}$ \\ t = 8.99} & \makecell{4.992$^{***}$ \\ t = 8.74} & \makecell{3.860$^{***}$ \\ t = 6.60} \\
\midrule
Region [UK] & \makecell{0.444$^{***}$ \\ t = 5.36} & \makecell{0.490$^{***}$ \\ t = 5.45} & \makecell{0.275$^{**}$ \\ t = 2.56} & \makecell{0.826$^{***}$ \\ t = 7.51} \\
\midrule
Type-token ratio & \makecell{-0.656 \\ t = -1.73} & \makecell{-0.415 \\ t = -1.01} & \makecell{-1.753$^{***}$ \\ t = -3.57} & \makecell{-1.227$^{*}$ \\ t = -2.44} \\
\midrule
Avg. sentence length & \makecell{-0.014$^{**}$ \\ t = -3.13} & \makecell{-0.016$^{**}$ \\ t = -3.26} & \makecell{-0.019$^{**}$ \\ t = -3.31} & \makecell{-0.012$^{*}$ \\ t = -2.00} \\
\midrule
Flesch Reading Ease & \makecell{-0.016$^{***}$ \\ t = -3.83} & \makecell{-0.017$^{***}$ \\ t = -3.65} & \makecell{-0.022$^{***}$ \\ t = -3.92} & \makecell{-0.013$^{*}$ \\ t = -2.35} \\
\midrule
Hedging ratio & \makecell{0.127 \\ t = 1.24} & \makecell{0.140 \\ t = 1.27} & \makecell{0.080 \\ t = 0.61} & \makecell{0.176 \\ t = 1.30} \\
\midrule
Assertive ratio & \makecell{0.196 \\ t = 1.81} & \makecell{0.208 \\ t = 1.77} & \makecell{0.245 \\ t = 1.75} & \makecell{0.267 \\ t = 1.86} \\
\bottomrule
Adjusted R$^{2}$ & 0.233 & 0.221 & 0.170 & 0.294 \\ 
F Statistic (df = 6; 187) & 10.77$^{***}$ & 10.11$^{***}$  & 7.59$^{***}$ & 14.40$^{***}$ \\ 
\bottomrule \\[-1.8ex] 
\end{tabular}
\end{table*}

To assess whether linguistic properties explained variation in LLM-generated evaluations, we fitted multiple linear regression models on combined data predicting each job-relevant metric from region (India vs. UK) and a set of linguistic features: type-token ratio, average sentence length, Flesch Reading Ease, hedging ratio, and assertive ratio.

The results of the regression analysis can be found in Table \ref{tab:regression_summary}. 
The regression models for hireablity, positive impression, self-promotion and storytelling were all statistically significant.
Across all models, region consistently emerged as a significant predictor, with UK transcripts receiving higher scores than Indian transcripts, even after controlling for linguistic characteristics. The scores were $> 0.4$ points higher for hireablity, positive impression, and storytelling compared to Indian transcripts.

Among the linguistic predictors, both average sentence length and FRE, i.e., easier-to-read language, were consistently and negatively associated with all LLM scores. 
While these two features are partially correlated, since longer sentences reduce reading ease, they each contribute uniquely to model predictions. The negative coefficients suggest that transcripts with simpler, more readable language (i.e., higher Flesch scores) were rated less favorably, while those with longer sentences were also penalized.  
In other words, LLMs may reward sophisticated vocabulary use (with lower Flesch scores) while penalizing excessive sentence length.
This may penalize speakers from linguistic backgrounds where longer, more formal sentence constructions are common (e.g., Indian academic English), thus contributing to unintended cultural bias \cite{bender2019rule, inbookkumar, fleisig2024linguisticbiaschatgptlanguage}.

Interestingly, type-token ratio (a proxy for lexical diversity) was also significantly and negatively associated with self-promotion and storytelling. 
When clarity in articulating accomplishments and narrative coherence is crucial, excessive lexical diversity may not be favoured.
Rather than rewarding lexical variety alone, LLMs may favor responses that maintain coherence and relevance while using moderately sophisticated vocabulary. 
Taken together, these findings point to a nuanced preference for concise, semantically dense language, rather than for either simplicity or complexity in isolation.

\subsection{Effect of Identity Cues}
We tested whether identity cues of gender, region, and caste embedded in candidate names affected LLM-generated evaluations of Indian interview transcripts. Linear mixed-effects models were fitted for each job-relevant metric across 400 data points, with transcript ID included as a random effect.
Across all four models, neither region nor caste had a statistically significant effect on scores ($p > .10$), indicating minimal sensitivity to these identity dimensions.
Gender showed small but statistically significant effects for hireability ($p = .031$) and positive impression ($p = .034$). However, these differences were not significant in post-hoc pairwise comparisons with TukeyHSD \cite{tukeyComparingIndividualMeans1949} (e.g., $p = .275$ for hireability), suggesting the effects may not be robust. 
Gender effects were non-significant for self-promotion ($p = .149$) and storytelling ($p = .095$).

Overall, these findings suggest that LLMs did not consistently alter their evaluations based on identity cues in names alone. These findings suggest that LLMs, when stripped of broader socio-contextual inputs, may not rely heavily on stereotypical associations tied to names alone in controlled textual settings.

\section{Discussion}
This study examined how large language models (LLMs) evaluate job interview performance across cultural and identity contexts, using systematically controlled interview transcript data from Indian and UK job seekers. Our analysis yielded three core findings: (1) a consistent scoring disparity disadvantaging Indian transcripts, (2) systematic associations between linguistic features and LLM evaluations, and (3) minimal effects of identity-based name cues in controlled settings.
Below, we discuss these findings and their broader implications.

\subsection{Systematic Cultural Disparities and Consequences for Workforce}
Our cross-cultural comparison revealed that Indian transcripts received significantly lower scores than UK transcripts. 
These differences persisted across two popular LLMs and were robust to anonymization of identity-related content.
Our findings suggest that LLM-based interview scoring can inadvertently favour Western linguistic patterns and communication norms (e.g. phrasing, tone), leading to systematic disadvantages for non-Western candidates. 
The immediate concern is that such biases could translate into less diverse hiring outcomes. If organizations rely on these AI evaluations for early-stage screening, qualified candidates who don’t fit the model’s learned linguistic mold may be filtered out, perpetuating the underrepresentation of non-Western groups in global hiring pipelines. Our findings expand on the previous studies showing favourable outcomes for Western contexts \cite{wilsonGenderRaceIntersectional2025} where an LLM-based resume screener significantly favoured White-associated names while consistently disadvantaging Black male candidates. 
In effect, ``hidden'' talent from marginalized communities can remain hidden \cite{fullerHiddenWorkersUntapped2021} if AI tools systematically overlook or undervalue non-normative styles and experiences. 
At scale, these exclusions reinforce global inequalities, risk legal and reputational harm for organizations, and replicate historical patterns of discrimination.
These structural biases not only raise ethical and legal red flags but also threaten to create a workforce that is homogeneous and missing perspectives from broad segments of society.

Furthermore, if these LLMs are used not only for evaluation but also for designing interviews \cite{raoImprovingAsynchronousInterview2021} or supporting interview training and preparation \cite{raoPotentialSupportingAutonomy2024, cherubiniElucidatingSkillsJob2021}, they risk introducing an additional layer of bias reinforcement.
AI-driven interview coaching tools, often trained on Western-centric data, may unintentionally pressure candidates to adopt Western communication styles, risking the loss of culturally rooted expression. While such tools can support skill development, they may penalize authenticity: candidates must either conform to dominant norms to improve their scores or maintain their natural style and risk lower evaluations.
Our results hinted at this effect: Indian candidates might need to modulate their natural speech patterns to score well under the LLM evaluator’s criteria.
More direct evidence comes from related studies of AI writing assistants. For example, \citet{agarwalAISuggestionsHomogenize2025} found that when a Western-centric LLM provided suggestions, Indian participants gradually adopted more Westernized writing styles, losing some of the linguistic nuances that reflected their cultural identity. 
Over time, such homogenization may suppress diverse ways of thinking and communicating in professional contexts. This highlights the need for AI coaching systems that value varied expressions of eloquence and competence, rather than reinforcing a singular cultural norm.

\subsection{AI Trust and Overreliance in Global Contexts}
Sociopsychological studies have found that individuals in more collectivist cultures often view AI as a helpful, authoritative extension of their community or self, whereas those in individualistic cultures are more likely to treat AI with skepticism, seeing it as a potential threat to personal autonomy \cite{barnesAICultureCulturally2024}. 
In practice, this means an HR manager in India or another collectivist context might place higher faith in an AI’s interview ratings, accepting them with little question, while a UK or US manager might be more inclined to double-check the machine’s judgments. If users assume the AI is neutral and objective, they may overlook subtle cultural prejudices influencing the scores. 
Overreliance is particularly dangerous when the model’s biases align with the user’s blind spots: for example, a Western-trained model and a non-Western recruiter both accepting Western norms as default. 

Introducing explainability into the AI’s evaluations offers one promising route to mitigate this risk. 
Research on human-AI decision-making has shown that transparent explanations can reduce overreliance on AI outputs by encouraging users to critically evaluate recommendations \cite{vasconcelosExplanationsCanReduce2023}. 
In the hiring context, if an LLM explains the reasoning behind a low score, such as pointing to specific communication traits or responses, a culturally aware recruiter can assess whether the feedback reflects a genuine job-relevant issue or a cultural mismatch. 
This transparency empowers human stakeholders to interrogate and override the AI’s judgment based on local norms and values.

More broadly, we argue that AI systems for hiring should be designed with intentional friction \cite{mejtoftDesignFriction2019}. While conventional UX principles emphasize seamless integration, reducing user effort, such smoothness can obscure the limitations and cultural assumptions embedded in AI evaluations. Instead, we advocate for a \textit{seamful design} \cite{ehsanSeamfulXAIOperationalizing2024} approach, which reveals the system’s boundaries, uncertainties, and imperfections to users. Rather than concealing these seams, designers can use them to prompt critical reflection and contextual adaptation. While Explainable AI has traditionally focused on illustrating algorithmic logic, seamful design can extend this by revealing sociotechnical and cultural mismatches, encouraging users to question not just how the AI works, but whether it aligns with their context and values.

\subsection{Toward Culturally Aware LLM Development}
Mitigating cultural harms requires rethinking how we develop and deploy LLMs in hiring.
Most AI bias and fairness evaluations to date have been Western-centric \cite{sambasivanReimaginingAlgorithmicFairness2021, septiandriWEIRDFAccTsHow2023, urmanWEIRDAuditsResearch2025}, often overlooking biases that matter in other regions. 
\citet{khandelwalIndianBhEDDatasetMeasuring2024} introduced the Indian-BhED dataset to evaluate caste and religion based stereotypes, dimensions largely absent from standard fairness benchmarks.
Their analysis revealed that popular LLMs frequently reproduced harmful stereotypes related to caste and religion, even more so than for attributes like gender or race that are commonly studied in the West. 

In contrast, our study did not observe significant differences in LLM evaluations with identity substitutions based on caste and geographic region within Indian context. 
This discrepancy may stem from the limited ability of LLMs to detect subtle cultural signals embedded in names alone, or from the highly controlled nature of our data, where anonymization and content uniformity may have masked stereotype effects. 
These findings underscore the importance of developing and applying localized benchmarks and evaluation frameworks that can capture culturally specific biases, embedding different contexts (beyond names) of underrepresented or marginalized populations.

Equally important is a community-centered evaluation strategy when assessing and improving model behavior. Rather than AI developers defining what constitutes “fair” or “acceptable” output, recent research demonstrates the value of involving the communities that the AI impacts \cite{qadriAIsRegimesRepresentation2023, ghoshGenerativeAIModels2024}. 
This can lead to a deeper understanding of AI’s blind spots and more culturally aware interventions.

Achieving equitable AI-driven hiring requires participatory approaches to LLM development that center global and culturally diverse perspectives. 
For developers of LLM-based hiring tools, this can include expanding training data to reflect varied dialects, communication styles and languages, and co-creating evaluation criteria with input from non-Western recruiters and job seekers. 
Model performance should be routinely audited using region-specific benchmarks (e.g., Indian-BhED) and informed by feedback from target user communities.
For HCI and AI ethics researchers, this involves engaging a broad set of stakeholders like job seekers, employers, resource groups, social scientists, and policymakers throughout the AI lifecycle to ensure community-centered design and evaluation.
On the regulatory front, standard bodies and policymakers can mandate cross-cultural bias impact assessments \cite{boguckaCodesigningAIImpact2024} for high-stakes AI systems. 
For instance, organizations could be required to audit AI interview platforms for disparate impact across cultural groups and provide transparent documentation on how language variation is handled. Requiring explainability in AI decisions, as discussed earlier, could be part of these standards to expose culturally misaligned reasoning. 

Ultimately, fostering culturally aware LLMs demands an expanded view of fairness, one that incorporates global voices and lived experiences \cite{sambasivanReimaginingAlgorithmicFairness2021}. By involving diverse communities in shaping AI systems, we not only improve fairness and accuracy but also affirm the agency and satisfaction of those navigating AI-mediated hiring. 

\subsection{Limitations and Future Work}
While this study provides important insights into the cultural dynamics of LLM-based hiring evaluations, some limitations warrant consideration.

\noindent
\textbf{ASR Limitations.}
The accuracy of interview transcripts was constrained by the use of an automatic speech recognition system (OpenAI Whisper), which can exhibit reduced accuracy for non-Western accents. A manual review of 30\% of transcripts showed most errors involved named entities rather than substantive content. As a corrective measure, we anonymized all named entities prior to analysis. However, subtle transcription errors may still have affected linguistic features, particularly in the Indian dataset. Using accent-aware ASR models or fully human-corrected transcripts could reduce these effects.

\noindent
\textbf{Participant and Data Limitations.}
The Indian and UK participant groups differed demographically: all Indian participants were university students aged 18–23, while the UK group spanned a broader age range, with half over 30 and potentially more professional experience. Though all participants from both groups were job seeking and preparing with self-paced interview preparations, this could have influenced their performance. Future studies should aim to match participants more closely on age or professional background to better isolate cultural effects from demographic influences.


To preserve privacy, all identifiable entities (other than names) universities, cities, and organizations, were anonymized. While necessary, this reduced context may have reduced the realism and richness of candidate responses, especially in sections where institutional affiliations or geographic ties contribute to stereotypes in self-presentation. 

\noindent
\textbf{LLM shortcomings.}
A potential source of bias lies in the prompt design itself. 
To minimize subjective framing, we structured the task around granular, literature-derived items for each metric. Nevertheless, prompt design decisions may still embed implicit author biases. Future studies could mitigate this through prompt co-design with domain experts or prompt randomization across LLM runs.

Finally, our results reflect the behavior of two specific LLMs (GPT-4o and Gemini Flash 1.5). As model architectures, training data, and fine-tuning strategies evolve rapidly, future research should test whether these patterns generalize across other models and under fine-tuning with culturally diverse and demographically balanced datasets.

\noindent
\textbf{Scope of the Study.}
We focused on generic behavioral interviews rather than technical or role-specific assessments to model early hiring stages where general attributes such as communication and self-presentation are most relevant. Future studies should extend this framework to later stages to examine whether cultural biases persist in specialized contexts.

We did not establish a human ground-truth baseline, making it possible that some score differences reflect genuine performance variation. Comparative evaluations by HR professionals across cultural contexts could calibrate LLM scores and quantify any penalties relative to human assessments. Nonetheless, given that both groups were randomly selected job seekers, the UK group’s consistently higher scores across all metrics remain concerning.

The study relied solely on transcribed text, excluding multimodal cues such as tone or body language that could amplify cultural differences in storytelling or self-promotion. Future work should incorporate multimodal inputs and embed identity cues beyond names.


Finally, our findings stem from a controlled evaluation setting; real-world hiring introduces further complexities, including recruiter persona, multi-stage processes, and AI–human interaction effects. Future work should assess how LLM scoring influences actual hiring outcomes and whether interventions such as explainable feedback improve fairness and trust.



\section{Adverse Impact Statement}

This study highlights the potential for large language models (LLMs) to systematically disadvantage candidates from non-Western cultural backgrounds in automated hiring evaluations. While our findings raise concerns about fairness, they are intended to inform the responsible development and use of LLMs in evaluative settings. Our experimental setup involved controlled comparisons and anonymized data to minimize harm to participants. Nevertheless, we acknowledge that releasing findings about bias could be misused to justify exclusionary hiring practices or to downplay the significance of other forms of bias not observed in our controlled settings. We strongly discourage such interpretations and emphasize that the goal of this work is to improve transparency, equity, and accountability in AI-driven hiring systems. We recommend that developers and decision-makers engage critically with our findings to reduce systemic harms and avoid reinforcing dominant cultural norms at scale.

\section{Ethical Considerations Statement}

All procedures in this study were approved by the appropriate institutional review board. Participants in the video interview data collection phase provided informed consent and were made aware of how their data would be used. To preserve privacy and mitigate identity-related harms, all transcripts were anonymized by removing or replacing identifiable information such as names, universities, and geographic locations. We took additional steps to ensure that LLM evaluations did not directly expose or amplify harmful stereotypes, including careful prompt engineering and the use of controlled identity substitutions. No deceptive practices were employed. We also note that all analyses were conducted post hoc on de-identified text data, and no hiring decisions were influenced by this research. The study’s goal is to surface risks and design recommendations that can contribute to more equitable and culturally inclusive AI systems.


\appendix
\label{sec:reference_examples}

\bibliography{aies2025,manual}

\end{document}